\documentclass[sigconf,screen]{acmart}

\usepackage{multirow}
\usepackage{color}
\usepackage{algorithm}
\usepackage{pdfpages}
\usepackage{algpseudocode}
\usepackage{subfigure}
\usepackage{xspace}

\newcommand{\codecomment}[1]{\hfill$\triangleright$ {#1}}

\newcommand{\approach}{DietCode\xspace}

\AtBeginDocument{%
  \providecommand\BibTeX{{%
    \normalfont B\kern-0.5em{\scshape i\kern-0.25em b}\kern-0.8em\TeX}}}

\copyrightyear{2022}
\acmYear{2022}
\setcopyright{acmcopyright}
\acmConference[ESEC/FSE '22]{Proceedings of the 30th ACM Joint European Software Engineering Conference and Symposium on the Foundations of Software Engineering}{November 14--18, 2022}{Singapore, Singapore}
\acmBooktitle{Proceedings of the 30th ACM Joint European Software Engineering Conference and Symposium on the Foundations of Software Engineering (ESEC/FSE '22), November 14--18, 2022, Singapore, Singapore}

\acmPrice{15.00}
\acmDOI{10.1145/3540250.3549094}
\acmISBN{978-1-4503-9413-0/22/11}

\begin{document}

\title{Diet Code Is Healthy: Simplifying Programs for Pre-trained Models of Code}


\author{Zhaowei Zhang$^1$, Hongyu Zhang$^2$, Beijun Shen$^1$, Xiaodong Gu$^{1}$}
\authornote{Xiaodong Gu is the corresponding author.}
\affiliation{%
  \institution{$^1$School of Software, Shanghai Jiao Tong University, China}
  \country{$^2$The University of Newcastle, Australia}
}
\email{andy_zhangzw@outlook.com}
\email{{bjshen,xiaodong.gu}@sjtu.edu.cn, hongyu.zhang@newcastle.edu.au}

\renewcommand{\shortauthors}{Zhang, et al.}

\begin{abstract}
 Pre-trained code representation models such as CodeBERT have demonstrated superior performance in a variety of software engineering tasks, yet they are often heavy in complexity, quadratically with the length of the input sequence.
 Our empirical analysis of CodeBERT's attention reveals that CodeBERT pays more attention to certain types of tokens and statements such as keywords and data-relevant statements. 
 Based on these findings, we propose \approach, which aims at lightweight leverage of large pre-trained models for source code. \approach simplifies the input program of CodeBERT with three strategies, namely, word dropout, frequency filtering, and an attention-based strategy that selects statements and tokens that receive the most attention weights during pre-training. Hence, it gives a substantial reduction in the computational cost without hampering the model performance. Experimental results on two downstream tasks show that \approach provides comparable results to CodeBERT with 40\% less computational cost in fine-tuning and testing. 
\end{abstract}



\settopmatter{printacmref=true} 

\maketitle

\section{Introduction}
Pre-trained models of code such as CodeBERT~\cite{feng2020codebert} have been the cutting-edge program representation technology, yielding remarkable performance on a variety of software engineering tasks such as code completion~\cite{ciniselli2021empirical}, code search~\cite{feng2020codebert}, and clone detection~\cite{lu2021codexglue}. Having pre-trained on large-scale code corpora, they demonstrate a better understanding of the semantics of source code than previous deep learning models such as code2vec~\cite{alon2019code2vec} and ASTNN~\cite{zhang2019astnn}.

Despite making a giant leap in accuracy, pre-trained models are often heavy in computation, which significantly hinders their applications in practice.
For example, the standard CodeBERT contains 125 million parameters and takes 28 hours to pre-train on 8.5M code. 
More critically, pre-trained models usually need to be fine-tuned before use, which is cost inefficient due to the large scale of parameters and training data. 
As such, it is highly desirable to identify the critical feature learned by pre-trained models and reduce the required computational cost by focusing only on the important information from model inputs~\cite{rabin2021simplification}. 

To understand critical information learned by pre-trained models, we conduct an empirical analysis of CodeBERT -- a pre-trained model for programming and natural languages. Our study aims to find out (i) what kinds of tokens CodeBERT pays the most attention to; and (ii) what kinds of statements are most important to CodeBERT when learning code representations. 
To answer these two questions, we categorize tokens and statements into a few classes and summarize the attention weights of each class that the pre-trained CodeBERT assigns to. 
Our results reveal that keywords and data types are the most critical tokens that CodeBERT focuses on. 
In terms of statements, CodeBERT pays more attention to \emph{method signatures} and \emph{return statements}, which show the overall functionality of a method. 

Based on these empirical findings, we propose \approach, a novel method that aims at lightweight leverage of large pre-trained models for source code. \approach reduces the computational complexity of pre-trained models by pruning unimportant tokens and statements from the input programs. 
Three pruning strategies are proposed, including word dropout, frequency filtering, and attention-based pruning.  
In particular, the attention-based pruning strategy selects tokens and statements that receive the highest attention from pre-trained models.
The program simplification algorithm formulates statement selection as a 0-1 knapsack problem where statements are regarded as items, and their attention weights are regarded as values. The algorithm selects statements (items) under the constraint of a given target length (capacity).

We apply \approach to two downstream tasks, namely, code search and code summarization. We measure the performance with relative length, FLOPs, and time cost, and compare our approach with baseline models, including the original pre-trained code model and SIVAND~\cite{rabin2021simplification}. 
Experimental results show that \approach provides comparable results as RoBERTa (code), with nearly 40\% less computational cost in fine-tuning and testing.  

Our contributions can be summarized as follows:
\begin{itemize}
    \item We conduct an in-depth empirical analysis of critical tokens and statements learned by CodeBERT. 
    \item We propose a novel program simplification approach for pre-trained programming language models that can significantly reduce the computation cost while retaining comparable performance.
    \item We extensively evaluate the proposed approach in two downstream tasks and show the effectiveness of our approach.
\end{itemize}

\section{Background}

\subsection{Pre-Trained Language Models}

Pre-trained language models such as BERT~\cite{devlin2018bert}, GPT-2~\cite{radford2019language}, and T5~\cite{raffel2020t5} have achieved remarkable success in a variety of NLP tasks~\cite{yang2019xlnet,clark2020electra,lample2019cross}. 
They refer to neural network models trained on large text corpora and can be fine-tuned to low-resource downstream tasks.

State-of-the-art pre-trained models are mainly built upon the Transformer architecture~\cite{vaswani2017attention}. 
The Transformer is a sequence-to-sequence learning model using the attention mechanism~\cite{bahdanau2015neural}. It is based on the encoder-decoder architecture, where a source sequence is encoded into hidden states and then taken as input to the decoder for generating a target sequence. Both the encoder and decoder contain multiple identical layers, and each layer consists of a multi-head self-attention network followed by a feed-forward network. Both of the outputs will be normalized before entering the next layer.

The key component of Transformer is the self-attention mechanism that represents a sequence by relating tokens in different positions~\cite{vaswani2017attention}. The goal of self-attention is to learn the important regions in the input sequence. Unlike traditional recurrent neural networks~\cite{cho2014encdec}, self-attention networks can learn the dependencies from distant tokens in parallel.
For an input sequence ($x_1$,\ldots, $x_n$) of length~$n$, the self-attention produces its representation ($z_1$, $z_2$ ... $z_n$) as
\begin{equation}
  \label{eq:attention}
    z_{i} = \sum_{j=1}^{n} \mathrm{Softmax}(\frac{(x_{i}W^Q) \cdot (x_{j}W^{K})^T }{\sqrt{d}}) \cdot x_jW^V
\end{equation}
where $W^Q$, $W^K$, and $W^V$ denote the model's parameters. $d$ is the dimension of the input matrix.
The model involves several attention heads, and each head's output is concatenated into the final result.

Pre-trained models usually involve large-scale parameters and consume huge computational resources. For example, BERT-base and BERT-large contain 110M and 340M parameters, respectively. As such, the lightweight leverage of pre-trained models is greatly desirable for research and practitioners~\cite{sanh2019distilbert}.

\subsection{CodeBERT}\label{CodeBERT}
Recently, researchers have adopted BERT for software engineering tasks and proposed CodeBERT~\cite{feng2020codebert}. CodeBERT is a bimodal pre-trained model for natural and programming languages, capturing semantic representations from programming languages~\cite{feng2020codebert}. The program representations learned by CodeBERT can be further utilized for downstream tasks such as code search and code summarization.

CodeBERT is built on top of a Transformer encoder~\cite{vaswani2017attention}. The optimization involves two tasks: masked language modeling (MLM) and replaced token detection (RTD). MLM 
masks two random tokens from the input pair of code and natural language comments, and aims to predict the original token in an extensive vocabulary. 
RTD involves two generators and a discriminator. The generators predict the original token for the masked token while the discriminator predicts whether the tokens are original or not.
After pre-training, CodeBERT can be adapted to downstream tasks through fine-tuning on the target dataset. 


\section{Empirical Analysis}

\subsection{Study Design}
In this section, we describe our study methodology and experimental setup. There are many levels of granularity for code, such as tokens, statements, and functions. As an in-depth study, we begin by investigating the atomic unit of source code, namely, tokens. We next investigate the statement-level knowledge learned by CodeBERT, which contains basic structures and semantic units.
We finally explore the function-level knowledge learned by CodeBERT through downstream tasks. 
In summary, we design our study methodology by addressing the following research questions:
\begin{itemize}
    \item \textbf{RQ1: What critical tokens does CodeBERT learn about?}
    We study the critical information CodeBERT learned at the token level by analyzing the attention weights assigned to these tokens and visualizing their relative importance. 
    \item \textbf{RQ2: What critical statements does CodeBERT learn about?}
    We further study the statements that CodeBERT assigns the most weights to. We classify code statements into common categories such as \emph{initialization}, \emph{assignment}, and \emph{return}, and present the attention weight of each category assigned by CodeBERT.
\end{itemize}

To answer these questions, we first need to know how to represent the key information learned by CodeBERT. In other words, how to measure the importance of each token and statement?
Since the heart of CodeBERT is the self-attention network, where hidden states of each token are calculated layer-by-layer according to the self-attention weights, we measure the importance of each token using the attention weights in the Transformer layers in CodeBERT after pre-training. 
Next, we will present the details of how to measure the importance of tokens and statements, respectively.

\subsubsection{Measuring Token Importance Using Attention Weights}
As introduced in Section~\ref{CodeBERT}, CodeBERT takes a sequence of source code tokens as input and produces a self-attention weight for each input token. Each weight measures how the corresponding token gains attention from other tokens in the input sequence. The higher the attention weights, the more attention that are paid by other tokens. 
Therefore, in our study, we measure the importance of each token using the attention weight.
CodeBERT has multiple self-attention layers and heads, each producing an attention weight for the same token. We average the attention weights of all layers and heads for each token. 
In our experiments, we go through the CodeSearchNet corpus~\cite{husain2019codesearchnet} and take as input each code snippet to the pre-trained CodeBERT. Then, we calculate the average attention weight that CodeBERT assigns to each token.


\subsubsection{Computing Attentions of Statements}
Having obtained the importance of each token, we compute the attention weight for each statement. Intuitively, the attention weight for a statement can simply be obtained using the average attention weights of all its tokens. However, different tokens in a statement have different importance. Without considering the global importance of each token, statements that consist of unimportant tokens could obtain even higher attention. Based on this concern, we regularize statement attention by penalizing unimportant tokens that gain fewer attention weights across the whole corpus. More specifically, we calculate the attention weight for a statement $S$ using a weighted average of attention weights of its tokens:
\begin{equation}
    a(S) = \sum_{t\in S} w(t)\cdot a(t)
\end{equation}
where $a(t)$ represents the attention weight of token~$t$ in statement~$S$; $w(t)$ denotes the normalized attention weight of token~$t$ in the whole corpus described in the previous section. The normalization is implemented using the Softmax function, namely,
\begin{equation}
    w(t) = \text{Softmax}_{t\in S}(a(t))
\end{equation}

Unlike tokens, statements are often unique in the corpus and cannot be included in a dictionary. 
To efficiently analyze statements, we classify statements into 21 categories such as \emph{method signature}, \emph{variable declaration}, and \emph{if condition}, and only present the average attention weight for each category. 
These categories are mainly guided by Java specification~\cite{gosling2000java}. 
In particular, the \emph{logger} category contains statements with standard logging functionalities such as \texttt{log4j}, \texttt{Logger}, and \texttt{println}.
Table~\ref{tab:statement_classification} shows the categories of statements we have summarized and the corresponding quantities in the CodeSearchNet corpus.
We can see that \emph{function invocation} and \emph{method signature} are the most common statements, while \emph{loop} statements such as \emph{while} are relatively rare.

\begin{table}[]
    \centering
    \small
    \caption{Statistics of statements. \emph{Arithmetic} means statements with only mathematical operations. \emph{Function Invocation} represents statements that invoke other functions.}
    \begin{tabular}{lc|lc}
    \toprule
    Category & Quantity & Category & Quantity \\
    \midrule
    Function Invocation & 16,558  & Throw & 1,460\\
    Method Signature & 11,755  & Catch & 1,309\\
    Variable Declaration & 11,701 & Arithmetic & 628\\
    If Condition & 11,646 & Case & 577\\
    Annotation & 8,980 & While & 459\\
    Return & 8,331 & Break & 341\\
    Getter & 3,092 & Finally & 297\\
    For & 2,190 & Continue & 142\\
    Try & 1,797 & Switch & 210\\
    Logging & 1,763 & Synchronized & 73\\
    Setter & 1,721 &  & \\
    \bottomrule
    \end{tabular}
    \label{tab:statement_classification}
\end{table}

\subsection{Results and Analysis}

In this section, we present the results of our experiments for each research question.

\subsubsection{What does CodeBERT learn about code tokens? (RQ1)}

\begin{figure}[h]
     \centering
     \includegraphics[scale=0.35]{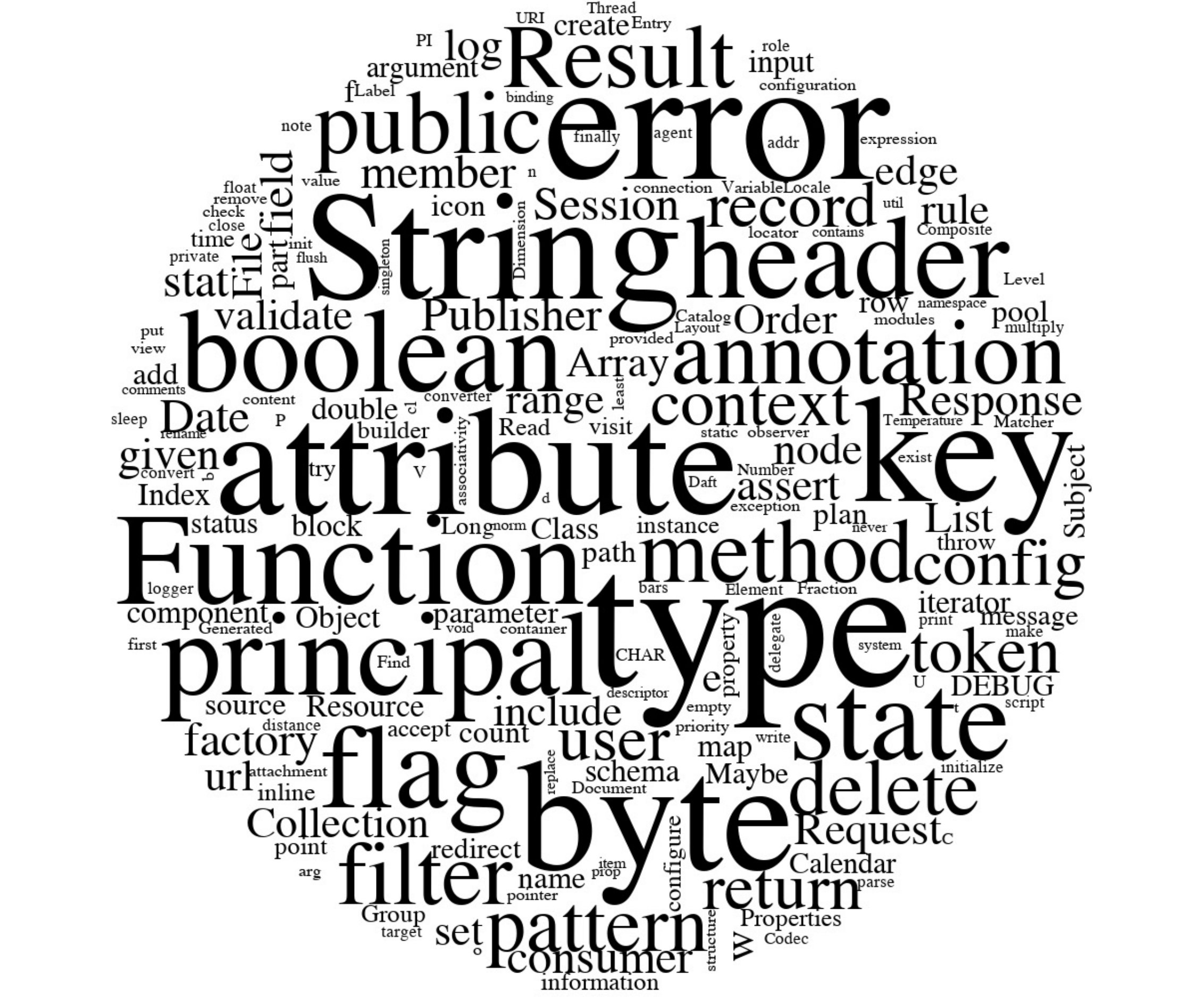}
     \caption{Java tokens highlighted by CodeBERT. The size of each token indicates the attention weight assigned by CodeBERT.}
     \label{fig:java_token_word_cloud}
 \end{figure}

Figure~\ref{fig:java_token_word_cloud} highlights the critical tokens in Java that gain the most attention by CodeBERT. The graph is summarized from the CodeSearchNet~\cite{husain2019codesearchnet} corpus which contains more than 100,000 Java functions. For tokens in each function, we calculate their attention weights and insert them into a <token, attention> map until the keys of the map are stable. 
For ease of visualization, we remove tokens that appear less than 50 times in the corpus. 
Among all the Java tokens, \texttt{coverage} gains the lowest attention weight (5.43$e$-5) while \texttt{boolean} has the highest weight (2.94$e$-2). The standard deviation of these weights is 5.97$e$-3.

The results show that CodeBERT assigns more attention to Java keywords such as \texttt{public} and \texttt{boolean}. This is expected since Java keywords frequently appear in Java code and play a dominant role in representing code semantics. Another category of tokens that receives high attention is the data-oriented identifiers such as \texttt{Map}, \texttt{List}, and \texttt{String}, probably because they define key data structures in a function.

\begin{figure*}
    \centering
    \includegraphics[scale=0.42]{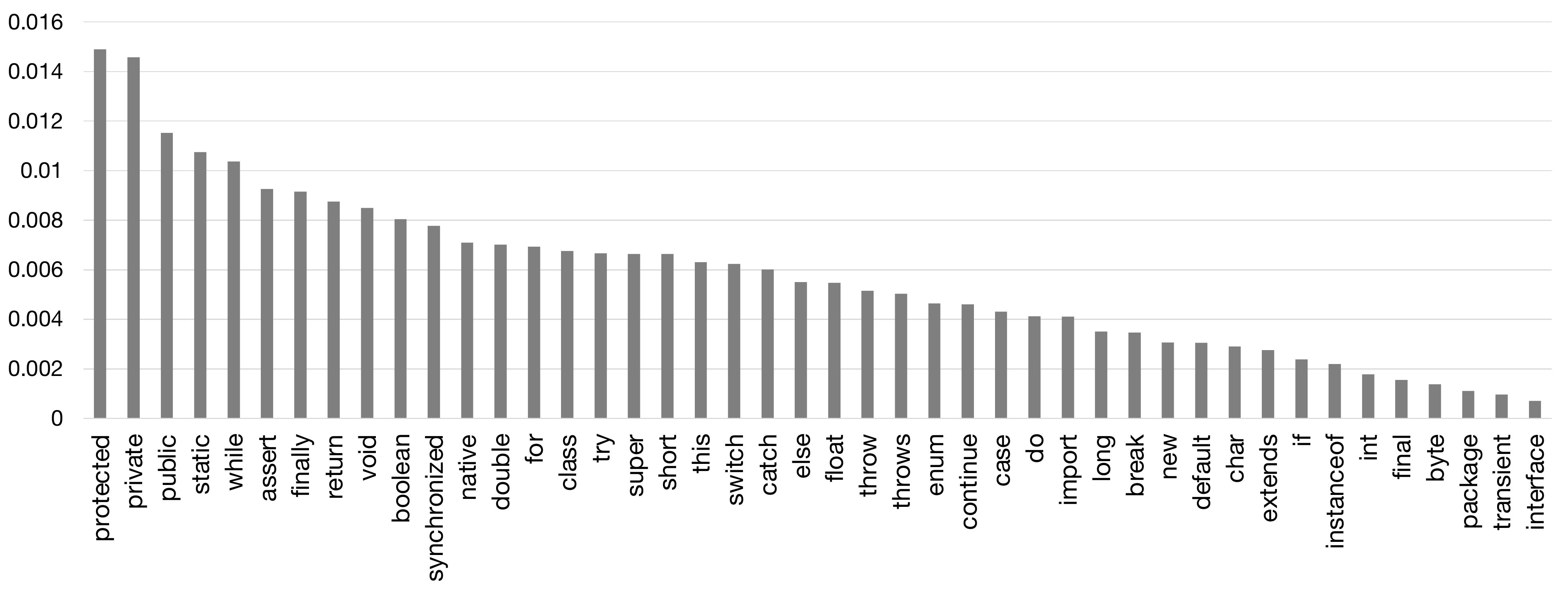}
    \caption{Attention weights of Java keywords learned by CodeBERT}
    \label{fig:java_keyword_attention}
\end{figure*}

\begin{figure}
    \centering
    \includegraphics[width=0.99\linewidth,height=0.8\linewidth]{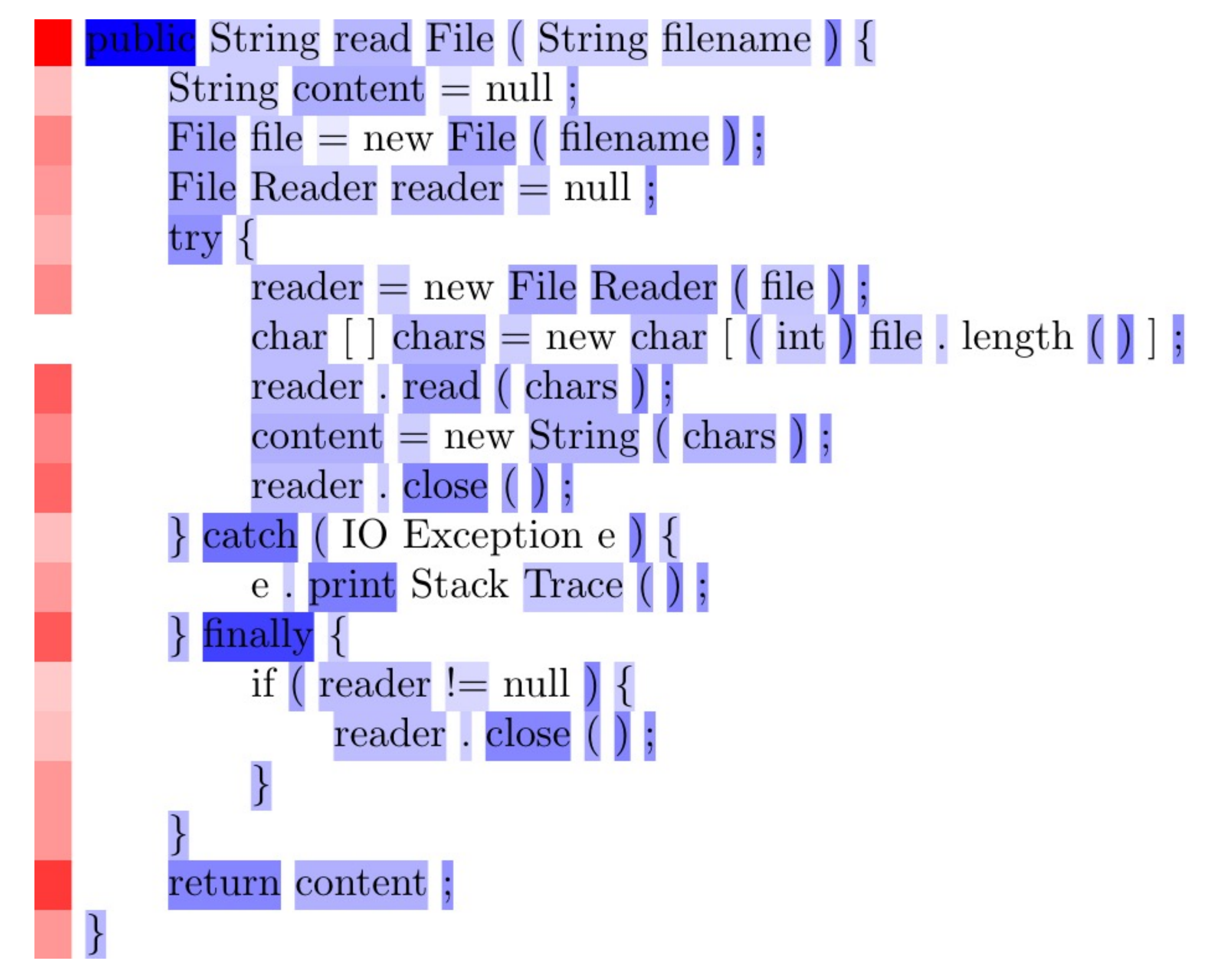}
    \caption{A heatmap of attention weights for Java statements and tokens. The background color of a token is proportional to the average attention weight assigned to it).}
    \label{fig:example_heatmap}
\end{figure}


Motivated by this finding, we further study the attention of Java keywords.
Figure~\ref{fig:java_keyword_attention} shows the attention weight of each Java keyword. As we can see, the keyword \texttt{private} obtains the highest attention weight of 1.15$e$-2 among all keywords, while \texttt{interface} and \texttt{transient} gain the lowest attention weights of 7.14$e$-4 and 9.62$e$-4, respectively. The standard deviation of these weights is 1.513$e$-3.


Among all Java keywords, method modifiers such as \texttt{public}, \texttt{private}, and \texttt{static} receive the highest attention weights 
We conjecture that these modifiers often denote the beginning of a method signature, which is essential to understanding the entire code. 
Next to these modifiers, \texttt{finally} and \texttt{return} also receive high attention weights.
\texttt{Finally} usually represents the end of a function, and the code inside the \texttt{finally} block will definitely be executed.
\texttt{Return} represents the method's output, which may show the method's functionality.
Surprisingly, branching-related keywords such as \texttt{if} and \texttt{switch} receive lower attention, presumably because CodeBERT considers them unimportant for program understanding despite their frequent occurrence in code. 

A similar pattern can be observed in Python. For example, CodeBERT prefers keywords such as \texttt{return} and \texttt{def} that represent the general structure of a function, just like \texttt{return} and \texttt{public} in Java. In particular, among the branching and loop keywords, \texttt{while} ranks higher than \texttt{if} and \texttt{for}.

Figure~\ref{fig:example_heatmap} shows a heatmap of attentions for a Java function. The code is about reading content from a file. The tokens with higher attention weights are marked in deeper colors. 
The heatmap shows a similar result as discussed above. \texttt{public} and \texttt{finally} receive the highest attention weights in this function, while \texttt{new} and \texttt{if} obtain the lowest attentions. 
Tokens that are related to the core functionality, such as \texttt{File} and \texttt{read}, have shown high attention weights. 
By contrast, tokens that are auxiliary to the main functionality, such as \texttt{null} and \texttt{Exception} have received much lower attention.
Another interesting observation is that grammatical symbols such as \} and; are considered necessary by CodeBERT probably because they mark the end of statements and blocks.
Overall, the heatmap shows that CodeBERT mainly pays attention to functional tokens that reflect the overall functionality and does not care much about notional tokens about grammar and special branches.

\subsubsection{What does CodeBERT learn about code statements? (RQ2)}

\begin{figure}
    \centering
    \includegraphics[width=0.95\linewidth,height=0.5\linewidth]{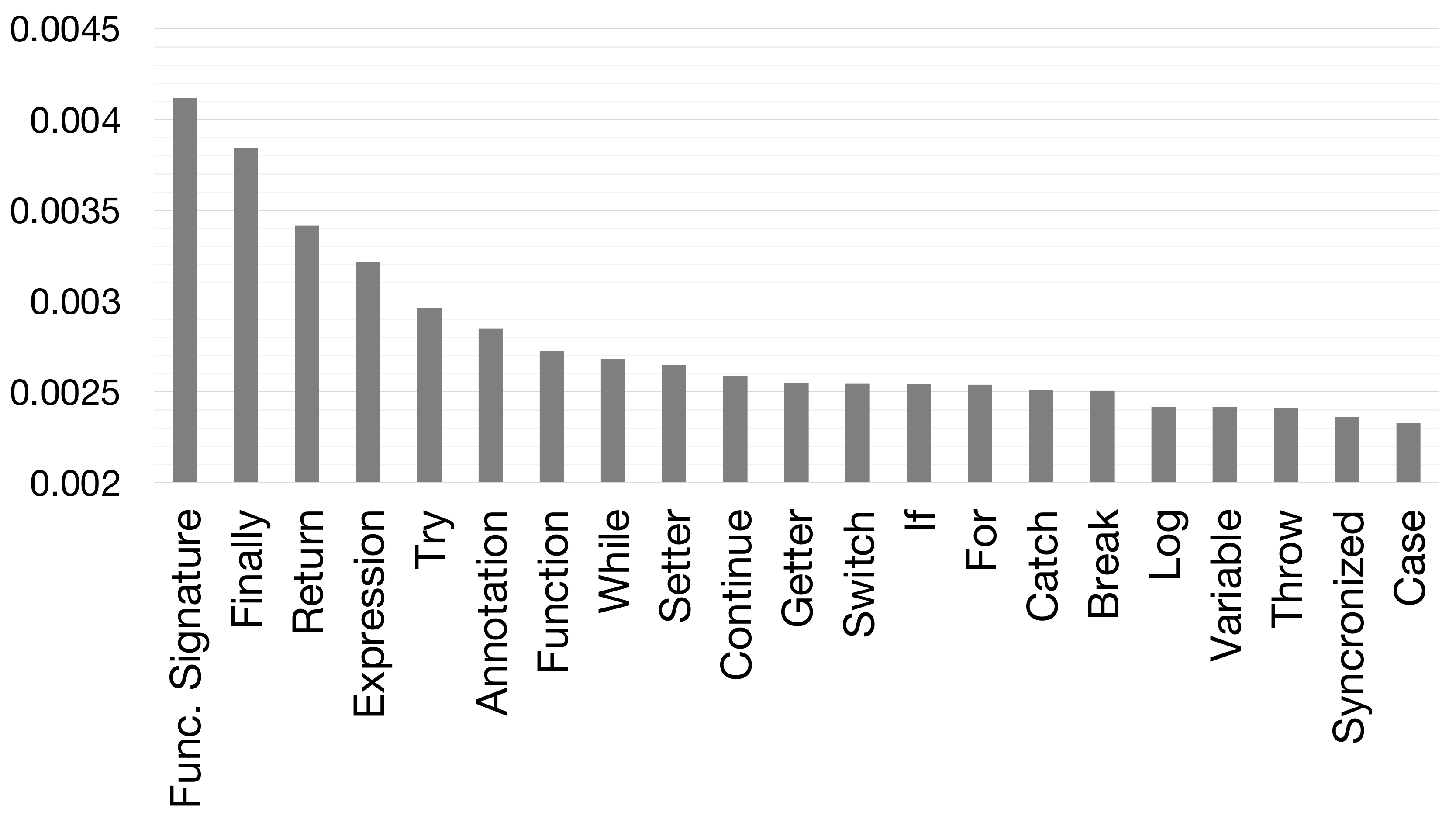}
    \caption{Attention weights of various types of Java statements learned by CodeBERT}
    \label{fig:java_statement_classification_attention}
\end{figure}

Figure~\ref{fig:java_statement_classification_attention} shows the attention weights assigned by CodeBERT for each type of Java statement. As seen, \emph{method\;signature} has gained the highest attention (4.11$e$-3), while \emph{case\;statement} obtains the lowest attention (2.32$e$-3). The standard deviation of the attention weights over all categories is 4.88$e$-4.
Broadly speaking, CodeBERT focuses more on statements indicating the overall functionality of a method, such as \emph{method signature} and \emph{return}, probably because they contain dense information of the whole function, such as names and targets.

Interestingly, \emph{arithmetic expressions} (expressions with mathematical operations)
also receive more attention than other statements. Like natural language, \emph{arithmetic expression} can be viewed as a sequence of operations that tell the functionality of a statement.
\emph{Function invocations} has also been shown to be essential for CodeBERT to understand code.
Like \emph{arithmetic expressions}, \emph{function invocations} can be regarded as identifiers of other method signatures, which can be literally understood through the function names and parameters.

Surprisingly, CodeBERT does not pay much attention to statements related to control flow structures, such as \emph{while}, \emph{for} and \emph{case}. 
This indicates that CodeBERT can effectively learn representations of plain texts while being limited in learning structures.

Figure~\ref{fig:example_heatmap} also shows the heatmap of statement attentions for a sample function of Java. Similar to the conclusions above, \emph{method signature} is the most important, which provides a general introduction to the method's functionality.
At the same time, statements about \emph{variable declaration and initialization}, such as ``\texttt{String content == null}'' receive lower attentions, probably because they only create new variables with initial value \texttt{null} without actual operations to them.
The attention weight of the \emph{if} statement in the \texttt{finally} block is also low (2.92$e$-3). This function only closes the file reader and may have little effect on CodeBERT's understanding of the whole method.

We observe a similar trend for Python statements. \emph{Method signatures} also receive the highest attention weight (6.86$e$-3), while branching statements such as \emph{break} and \emph{continue} have the lowest weights. 
The standard deviation of attention weights for Python statements is 1.66$e$-3, which is a bit larger than that of Java statements. 

Compared to Java, the distributions of Python statements and tokens are a bit more sparse. For example, the \emph{method signature} statement which ranks the first owns a much higher weight than the second category of statements (i.e., \emph{return}). The top keyword \texttt{def} also obtains a much higher attention weight than other keywords. 



\section{\approach: Program Simplification for CodeBERT}
As Equation~\ref{eq:attention} indicates, the computational complexity for computing the attention matrix is $\mathcal{O}(d^2n + n^2d)$, which quadratically scales with sequence length. As such, the attention operation becomes a bottleneck when applied to long sequences such as source code~\cite{kim2021learned}.

We wonder whether we can remove unimportant tokens or statements from the input programs of CodeBERT. In this way, the time and space costs for the pre-trained model can be substantially reduced.
Based on this idea, we propose \approach, a lightweight pre-trained model for source code. \approach reduces the computational complexity of CodeBERT by simplifying input programs. 

More specifically, our problem formulation is as follows: given a code snippet $C=\{s_1;\ldots;s_{|C|}\}$ which  consists of $|C|$ statements, we want to output a pruned code snippet $C_p$ which contains a maximum length of $L$ tokens. The goal of program simplification is to prune as many tokens as possible without a significant impact on the accuracy of downstream tasks. 


\subsection{Word Dropout}
One straightforward strategy for program simplification is the word dropout~\cite{iyyer2015deep}, namely, randomly dropping tokens from the input code to meet a specified sequence length. 
For each token $w\in S$, we define a binary variable~$r_w\in\{0, 1\}$ to indicate whether the token will be kept ($r_w=1$) or pruned ($r_w=0$):
\begin{equation}
  \begin{split}
    & r_w \sim \mathrm{Bernoulli} (p)\\
    & C_p = \{w|w\in C~~\mathrm{and}~~r_w>0\} 
  \end{split}
\end{equation}
where $p=L/|C|$ denotes the probability of selecting a token. 
In addition to reducing computational complexity, word dropout has also been shown to improve the robustness of neural networks~\cite{iyyer2015deep}. Hence, it enhances the performance of many tasks.

\subsection{Frequency Filtering}
A key concern with the word dropout strategy is that it can also prune important tokens randomly from the input. 
Hence, we introduce another frequency-based token pruning strategy that removes uncommon tokens while keeping only the most frequently occurred tokens. 

Specifically, for each input code snippet $C = \{w_1,...,w_n\}$, we keep tokens if their frequency is among the top $k$ ($k=|C_p|$) of all tokens in the input, namely,
\begin{equation}
    C_p = \{w|w\in C, f(w)\in\mathrm{top-}k(f(w_1),...,f(w_n)), k=|C_p|)
\end{equation}
where $f(w)$ denotes the frequency of $w$ in the whole code corpora.

\subsection{Attention-based Code Pruning}

The frequency filtering strategy can roughly distinguish important tokens, but may only bias common tokens. According to our empirical findings above, CodeBERT pays attention to certain types of tokens and statements. Tokens that receive high attention do not coincide with common tokens. 
In order to provide a more fine-grained selection of important tokens and statements, we propose an attention-based code pruning strategy that selects tokens and statements based on their attention weights.

\begin{algorithm}
    \caption{The Attention-based Pruning Algorithm}
    \begin{algorithmic}[1]
     \Require \\
     \texttt{C}=$s_1;\ldots;s_{|C|}$: an input code snippet, \\
     $\mathbf{A}^T$: attention dictionary of tokens,\\
     $\mathbf{A}^S$: attention dictionary of statements,\\
     L: target length
     \Ensure \\
     Reduced code snippet $\texttt{C}_p = s'_1;\ldots;s'_{|C_p|}$
    \For {s $\in$ \texttt{C}}
        \State a(s) = $\mathbf{A}^S$(s) \codecomment{obtain the statement attention}
        \For {t $\in$ s}
            \State a(t) = $\mathbf{A}^T$(t) \codecomment{obtain the token attention} 
        \EndFor
    \EndFor
    \State $C_0$ $\leftarrow$ \texttt{0-1 Knapsack} (items=\{s\}$_{s\in C}$, values=$\{a(s)\}_{s\in C}$, weights= \{|s|\}$_{s\in C}$, capacity = $L$+$\text{max}_{s\in C}$(|s|) \codecomment{candidate statements}
    \For {i = 1,..., $\sum_{s\in C_0}$(|s|) - L}
        \State $s'$ $\leftarrow$ $s'$ $\setminus$ \{t\} \quad where $s'=\mathrm{argmin}_{s \in C_0}a(s)$,\newline 
        \phantom{sp~ $s'$ $\leftarrow$ $s'$ $\setminus$ \{t\} \quad where {}} $t=\mathrm{argmin}_{t_0\in s'} a(t_0)$ 
    \EndFor
    \State $C_p$ $\leftarrow$ $s'_1;\ldots;s'_{|C_0|}$\quad where $C_0$ = $\{s'_1, s'_2,...,s'_{|C_0|}\}$ \codecomment{concatenate statements}
    \end{algorithmic}
    \label{algo:prune}
\end{algorithm}

The simplification procedure is summarized in Algorithm~\ref{algo:prune}. 
We begin by summarizing the category of statements for a given programming language according to the methodology in our empirical study. Then, we create attention dictionaries for both statements and tokens.
The algorithm runs code pruning in two phases: \emph{selecting statements} and \emph{pruning tokens}.

In the statement selection phase, we select pivot statements that belong to a category that obtain the highest attention weights in our empirical study. Meanwhile, we want the selected statements not to exceed the length constraint (i.e., $L$ tokens).
This can be formulated as a 0-1 knapsack problem~\cite{pisinger1995algorithms}, where statements can be regarded as the items to be collected into a knapsack, with the attention weights being the values, and the statement lengths being the weights. The length constraint can be regarded as the capacity of the knapsack. 
It is worth noting that we enlarge the capacity by the maximum length of statements for tolerance of selecting one more statement so that the algorithm can further perform token pruning in the next phase.   

In the token pruning phase, we greedily remove the tokens that have the lowest attention weights in the lowest weighed statements iteratively until satisfying the maximum number of tokens. 

One potential problem is the different scales of statement attentions and token sizes. 
We find that some shorter sequences are much more likely to be chosen during our experiments even if they receive low attention weights.
This is probably because the value ratio (the attention weight divided by the sequence length) of a shorter sequence is much larger than the longer one.
We amplify the attention weights using min-max normalization and multiply with the number of tokens in the statement, that is,
\begin{equation}\label{attention}
    a(s) = \frac{a(s) - \mathrm{min}_{s\in S}(a(s))}{\mathrm{max}_{s\in S}(a(s)) - \mathrm{min}_{s\in S}(a(s))}\times N
\end{equation}
where $N$ denotes the number of tokens in the statement.

The time complexity of the simplification algorithm is $\mathcal{O}$($N \cdot (L_{T} + L_{M})$), where $L_{T}$ denotes the target number of tokens, and $L_{M}$ denotes the maximum length of statements. The time cost mainly comes from the 0-1 knapsack algorithm. Furthermore, the knapsack algorithm requires a two-dimensional array for dynamic programming, which requires space complexity.

\begin{figure*}
    \subfigure[The original source code]{
            \fbox{\includegraphics[scale=0.33]{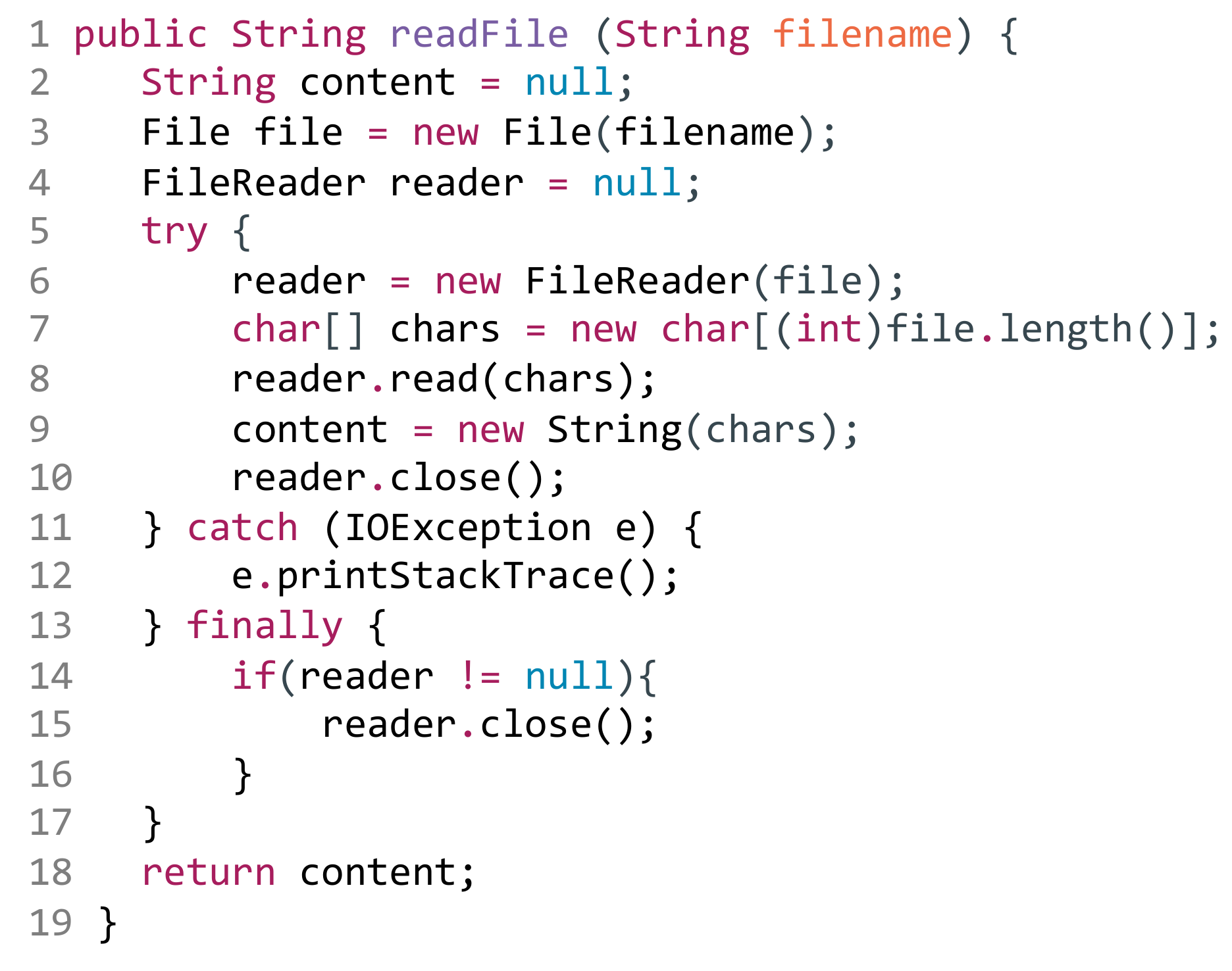}}
    }
    \subfigure[Code after statement selection]
    { 
        \fbox{\includegraphics[scale=0.33]{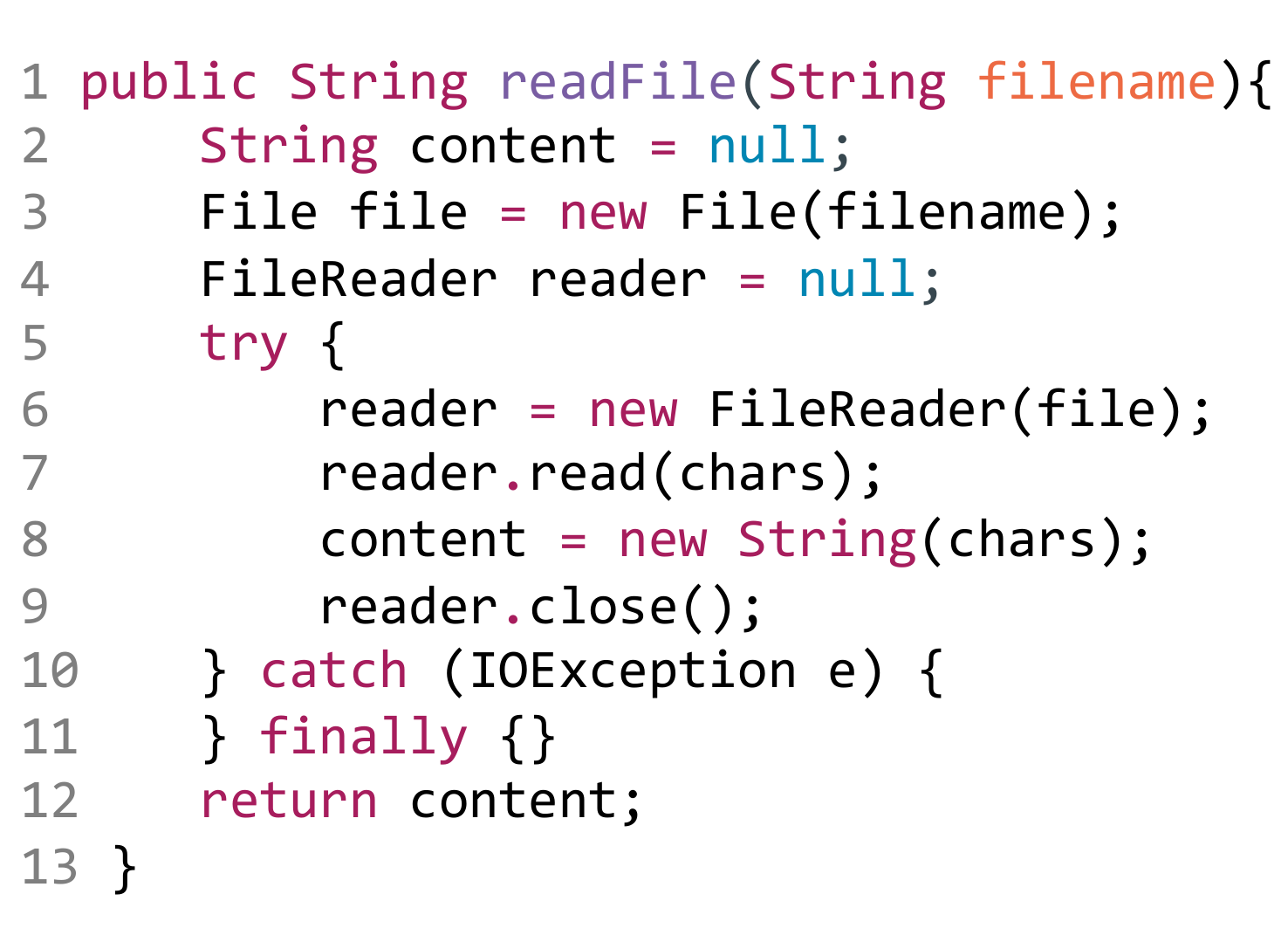}}
    }
    \subfigure[Code after token pruning]
    { 
        \fbox{\includegraphics[scale=0.33]{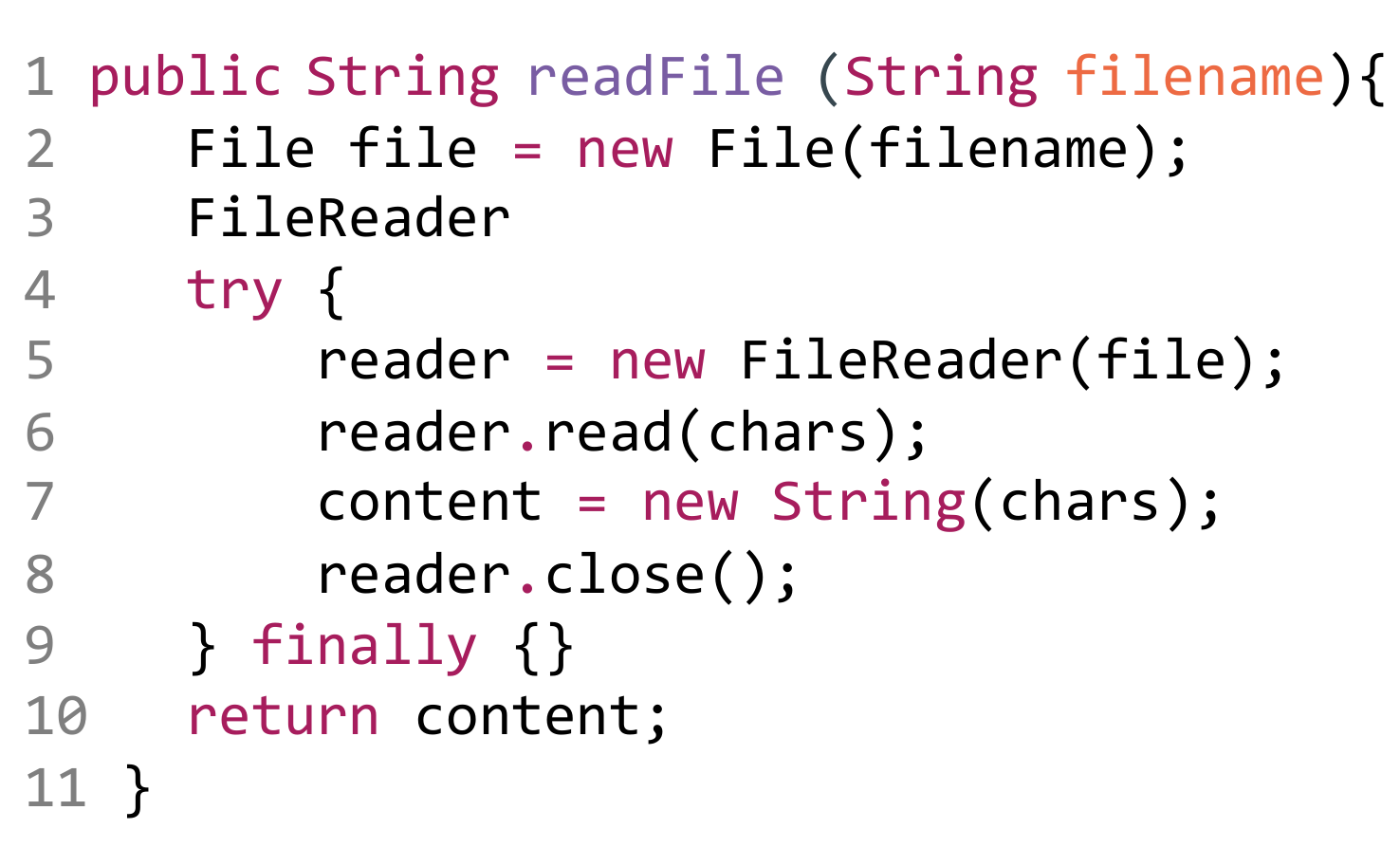}}
    }
    \caption{An example of program simplification by \approach~(attention).}
    \label{fig:code_example}
\end{figure*}

Figure~\ref{fig:code_example} shows an example of program simplification by \approach. The original code (Figure~\ref{fig:code_example}(a)) aims to read content from a file with a given filename. 
Our goal is to reduce the code to contain only 60 tokens without considerably losing the semantics.
In the \emph{statement selection} phase (Figure~\ref{fig:code_example}(b)), the simplification algorithm solves a 0-1 knapsack problem by setting the capacity to 78 (including the target length of 60 and the longest statement length 18). Then, \approach selects trunk statements such as \emph{method signature} and \emph{return}, and removes trivial statements such as \emph{variable assignments} and \emph{catch}.
The number of tokens in the code is reduced from 118 to 77.
In the \emph{token pruning} phase, \approach further removes tokens such as variable initialization (Figure~\ref{fig:code_example}(c)), and reduces the number of tokens from 77 to 60.
Although reduced by around 50\%, the main semantic of this function is not destroyed significantly. CodeBERT can still understand the function through key statements such as the method name~\texttt{readFile}, the key variable~\texttt{content}, and the invoked function~\texttt{read}. Although the remaining code is never runnable, the selected critical information can be used for downstream tasks such as code search and summarization.

\section{Evaluation}

This section introduces the evaluation of \approach in two downstream tasks.
We aim to answer the following research questions through experiments:

\begin{itemize}
    \item \textbf{RQ3: How effective is \approach in program simplification?} We evaluate the performance and computation cost of \approach in two downstream tasks and compare it with baseline models. 
    \item \textbf{RQ4: How effective is \approach under different relative lengths?} We study the effect of different relative lengths on the evaluation scores. Our goal is to find the relative length that leads to the best trade-off between model performance and computational expense.
    \item \textbf{RQ5: What is the effect of different pruning strategies?}  The attention-based \approach performs both statement and token pruning. We conduct an ablation study to identify the effect of pruning each of them. 
\end{itemize}

\subsection{Downstream Tasks}
We test our algorithm in two downstream tasks, namely, code search and code summarization. They are the most widely used software engineering tasks to demonstrate the capacity of NL-PL understanding~\cite{kamiya2002ccfinder, white2016deep, haiduc2010supporting, jiang2017automatically, nie2016query, stolee2014solving}.  

\emph{Code Search.}\;
Code search is a typical testbed for pre-trained code models such as CodeBERT~\cite{feng2020codebert}. The task aims to find code snippets from a codebase that is most relevant to a given query. 



\emph{Code Summarization.}\;
Code summarization aims to generate a natural language summary for an input code snippet. It has also been widely used for evaluating pre-trained models for source code~\cite{feng2020codebert,wang2021codet5}.
Through this task, we want to verify whether program simplification by \approach is effective in generation tasks.



\subsection{Datasets}

\begin{table}
    \caption{Statistics of Datasets}
    \centering
    \begin{tabular}{lcc}
    \toprule
    Corpus & Training & Test\\
    \midrule
    CodeSearchNet (Java) & 908,886 & 1,000,000\\
    CodeSearchNet (Python) & 824,343 & 1,000,000\\
    \bottomrule
    \end{tabular}
    \label{tab:dataset}
\end{table}

We use the CodeSearchNet to fine-tune and test our model~\cite{husain2019codesearchnet}, which contains <comment, code> pairs collected from open source projects.
\emph{code} means the code snippet of a method, while \emph{comment} is the description of the code that was mainly collected from comments for the whole functions such as Javadocs or docstrings in Python. 
The statistics of CodeSearchNet is shown in Table~\ref{tab:dataset}.

The original data of CodeSearchNet are in the format of token sequences. We preprocess the data by parsing them into statements.
We parse statements by splitting brackets and semicolons according to the common guidance of Java~\cite{gosling2000java}. 
For Python code, there are no delimiters because the indentation is missing in the dataset. Besides, Python does not use brackets and semicolons. Hence, we split statements using other hint symbols such as ``def'', ``='', ``:'', and ``)''.

To reduce the vocabulary size of code tokens, we delexicalize all string constants to a generic token \texttt{string} and all numerical constants to the same token \texttt{10}. We conserve special numbers such as 0, 1, and -1, which can belong to boolean values.

\subsection{Evaluation Metrics}

First, we want to measure the maximum extent of code reduction that \approach can achieve without losing much accuracy. We define \textbf{Relative Length (RL)} 
to measure how much of the code is remained after simplification. It is computed as the percentage of the length of the simplified code snippet $|C_p|$ with respect to the length of the original code $|C|$: 
\begin{equation}
    RL = \frac{|C_p|}{|C|}\times 100\%
\end{equation}
where |$\cdot$| denotes the length of a code snippet (i.e., number of tokens).
The smaller the relative length, the greater the simplification to the original code. 

We also use \textbf{FLOPs} (floating point operations)~\cite{hunger2005floating} to measure the effect of model reduction. FLOPs is a widely used metric to measure the complexity of a machine learning model. The higher the FLOPs, the slower the processing of the model. 

The third metric is the time cost, including \textbf{fine-tuning time (FT Time) and testing time}.
We calculate the execution time of \approach for fine-tuning and testing, measured by the number of hours from the program starting to the termination.

Besides the metrics for complexity, we use two standard metrics to evaluate the effectiveness of \approach in two downstream tasks respectively. 
We measure the performance of code search using \textbf{MRR} (mean reciprocal rank), which refers to the average of the multiplicative inverse of the position for the first correct answer for the query~\cite{husain2019codesearchnet}.

We measure the performance of code summarization using the \textbf{BLEU-4} (bilingual evaluation understudy) score~\cite{lewis2020bart}, which calculates the average of n-gram precision on a couple of sequences with a penalty for short sequences.

\begin{table*}
    \centering
    \small
    \caption{Performance of various code simplification methods on the code search task ($L_\mathrm{in}$=input length; RL=relative length; FT=fine-tuning).}
    
    \begin{tabular}{lccc@{}cclcccc@{}ccc}
    \toprule
     & \multicolumn{6}{c}{Java}& & \multicolumn{6}{c}{Python} \\ \cline{2-7}\cline{9-14}
     Model & $L_\mathrm{in}$ & RL & FT Time\;\; & Test Time  & FLOPs & MRR & & $L_\mathrm{in}$ & RL & FT Time\;\; & Test Time  & FLOPs & MRR\\
    \midrule
        BiRNN~\cite{cho2014encdec} & 200 & 100\% & 9.88h & 1.13h & 8.34G & 0.29 & & 200 &100\% &7.07h &1.25h & 8.34G & 0.32 \\
    SelfAttn~\cite{vaswani2017attention} & 200 & 100\% & 22.83h & 4.00h & 16.99G & 0.59 & & 200 & 100\%&17.78h & 2.67h &16.99G & 0.69  \\ 
        RoBERTa~\cite{liu2019roberta} & 200 & 100\%& 23.85h & 4.78h &16.99G & 0.67 & &200 &100\% & 19.32h & 3.53h &16.99G & 0.81   \\ 
    RoBERTa (code) & 200 & 100\%& 21.02h & 4.53h & 16.99G & 0.72& &200 &100\% & 20.78h & 4.07h & 16.99G &  0.84  \\ 
    \hline
        \textbf{CodeBERT}~\cite{feng2020codebert}  & 200 & 100\% & 20.82h & 3.17h & 16.99G & 0.74 & & 200 & 100\%   & 17.92h & 2.82h & 16.99G & 0.84\\ 
        \bf \approach &&&&&&&&&&&&& \\
        - Attention & 120 & 60\% & 11.08h & 1.83h  & 10.19G& 0.71& & 120& 60\%   & 9.62h & 1.82h & 10.19G & 0.81 \\
        - Dropout & 120 & 60\% & 10.73h & 1.95h & 10.19G & 0.68 & & 120 & 60\% & 9.33h & 1.93h & 10.19G & 0.80 \\
        - Frequency & 120 & 60\% & 10.32h & 1.8h & 10.19G & 0.66 & & 120 & 60\% & 8.67h & 1.79h & 10.19G & 0.78\\
    \midrule
    \textbf{CodeT5}~\cite{wang2021codet5} & 200 & 100\% & 16.83h & 2.97h & 16.99G & 0.72 &  & 200 & 100\% & 17.61h & 2.83h & 16.99G & 0.837 \\ 
    \bf \approach &&&&&&&&&&&&&\\
    - Attention & 120 & 60\% & 9.3h & 1.74h & 10.19G & 0.71 & & 120 & 60\% & 8.31h & 1.81h & 10.19G & 0.813\\
    - Dropout & 120 & 60\% & 9.25h & 1.67h & 10.19G & 0.68 & & 120 & 60\% & 9.33h & 1.93h & 10.19G & 0.799\\
    - Frequency & 120 & 60\% & 8.97h & 1.58h & 10.19G & 0.66 & & 120 & 60\% & 8.67h & 1.79h & 10.19G & 0.785\\
    
    \bottomrule
    \end{tabular}
    \label{tab:codesearch_result}
\end{table*}

\begin{table*}
    \centering
    \small
    \caption{Performance of various code simplification methods on the code summarization task.}  
    \begin{tabular}{lcc@{}c@{}c@{}ccccc@{}c@{}c@{}cc}
    \toprule
     & \multicolumn{6}{c}{Java}& & \multicolumn{6}{c}{Python} \\ \cline{2-7}\cline{9-14}
     Model & $L_\mathrm{in}$ & RL\;\;\; & FT Time\;\; & Test Time\;\;  & FLOPs & BLEU-4 & & $L_\mathrm{in}$ & RL\;\; & FT Time\;\; & Test Time\;\;  & FLOPs & BLEU-4\\
    \midrule
    Seq2Seq~\cite{cho2014encdec} & 256& 100\% & 6.26h & 1.26h & 12.750G & 15.09 & & 256&100\% & 4.43h & 1.02h & 12.750G & 15.93  \\ 
    Transformer~\cite{vaswani2017attention} & 256&100\% & 13.28h & 5.43h & 24.328G & 16.26 & &256 &100\% & 9.70h & 2.43h &24.328G & 15.81  \\  
    RoBERTa~\cite{liu2019roberta} & 256&100\% & 15.42h & 5.22h & 24.328G& 16.47 & &256 &100\% & 10.63h & 1.88h & 24.328G&   18.14\\ 
    RoBERTa (code) & 256&100\% & 14.5h & 6.02h & 24.328G& 17.50 & &256 &100\% & 10.45h & 2.05h & 24.328G& 18.58 \\ 
    \hline
    \textbf{CodeBERT}~\cite{feng2020codebert} & 256 & 100\% & 13.82h & 4.8h  & 24.328G& 18.95 & & 256 & 100\%   & 8.32h & 1.87h  & 24.328G& 19.04\\ 
    \bf \approach&&&&&&&&&&&&&\\
        - Attention & 150 & 60\% & 8.18h & 1.62h & 15.325G & 17.29 & & 150 & 60\% & 5.35h & 1.16h & 15.325G & 17.08 \\
        - Dropout & 150 & 60\% & 7.95h & 1.40h & 15.325G & 15.63 & & 150 & 60\% & 6.81h & 1.15h & 15.325G & 16.04\\
        - Frequency & 150 & 60\% & 8.62h & 1.33h & 15.325G & 16.13 & & 150 & 60\% & 6.12h & 1.42h & 15.325G & 16.55 \\
    \midrule
    \textbf{CodeT5}~\cite{wang2021codet5} & 256 & 100\% & 16.91h & 4.88h & 54.01G & 20.46 &  & 256 & 100\% & 9.68h & 1.48h & 54.01G & 20.37\\ 
    \bf \approach &&&&&&&&&&&&& \\
    - Attention & 150 & 60\% & 10.25h & 1.63h & 37.81G & 19.25 & & 150 & 60\% & 6.57h & 0.90h & 37.81G & 18.53 \\
    - Dropout & 150 & 60\% & 9.95h & 1.68h & 37.81G & 17.65 &  &150 & 60\% & 6.67h & 0.82h & 37.81G & 17.07\\
    - Frequency & 150 & 60\% & 10.23h & 1.35h & 37.81G & 18.74  &  & 150 & 60\% & 6.33h & 0.85h & 37.81G & 17.77\\
    \bottomrule

    \end{tabular}
    \label{tab:codesummarize_result}
\end{table*}


\vspace{-3pt}
\subsection{Experimental Setup}
The demonstrate the strength of our approach, we simplify the input programs for pre-training based models and compare the performance to that of the original pre-trained model. Specifically, we compare our technique to the vanilla versions of three pre-trained models:
\begin{itemize}
    \item \textbf{CodeBERT}~\cite{feng2020codebert}: the original CodeBERT without code simplification.
    We follow the experimental setup in the CodeBERT paper.
    \item \textbf{CodeT5}~\cite{wang2021codet5}: a widely used pre-trained programming language model based on the sequence-to-sequence architecture. CodeT5 has demonstrated better performance on generative tasks~\cite{wang2021codet5}.
    \item \textbf{RoBERTa}~\cite{liu2019roberta}: a popular extension of BERT that has demonstrated significant improvement. The original RoBERTa was pre-trained in natural languages. So we also report the results of RoBERTa~(code)~\cite{feng2020codebert}, a variant that was pre-trained on source code.
\end{itemize}

Besides the pre-trained models, we also compare the accuracy of \approach with that of traditional deep learning approaches, including \textbf{BiRNN}~\cite{cho2014encdec}, \textbf{SelfAttn}~\cite{vaswani2017attention}, \textbf{Seq2Seq}~\cite{cho2014encdec}, and \textbf{Transformer}~\cite{vaswani2017attention}. As they are also baseline models for CodeBERT, we directly take the results from the CodeBERT paper~\cite{feng2020codebert}.

Finally, we compare \approach with \textbf{SIVAND}~\cite{rabin2021simplification}, which is also a program simplification method proposed by \citet{rabin2021simplification}.
    SIVAND~\footnote{https://github.com/mdrafiqulrabin/SIVAND} is a machine-learning-based algorithm that recursively splits source code into fragments and measures its importance through the method name prediction task. The algorithm selects important fragments as the reduced code.
    In our experiments, we directly use the code provided by the authors.
    


We implement our model based on the GitHub repository of CodeBERT~\footnote{https://github.com/microsoft/CodeBERT} and CodeT5~\footnote{https://github.com/salesforce/CodeT5} using the default hyperparameter settings. All models are optimized with the Adam algorithm~\cite{kingma2017adam} with learning rates of 1$e$-5 and 5$e$-5 for the code search and code summarization tasks respectively.

We run all models on a machine with a CPU of Intel(R) Xeon(R) Silver 4214R 2.40GHz and a GPU of Nvidia Tesla P40.

\subsection{Experimental Results (RQ3)}

Table~\ref{tab:codesearch_result} and \ref{tab:codesummarize_result} show the performance of \approach in two downstream tasks. 
Overall, \approach reduces the computational cost by around 40\% while keeping a comparable accuracy to the original pre-trained models. 
The original CodeBERT takes nearly one day for fine-tuning and more than 3 hours for code search on the Java test set to achieve an MRR score of 0.74. 
In contrast, by reducing the length of the input code to 60\%, the fine-tuning and testing time can be reduced to 11.08 hours and 1.83 hours, respectively. The FLOPs also drops from 16.99G to 10.19G. At the same time, the accuracy (MRR = 0.71) is comparable to the original CodeBERT (MRR=0.74). It is worth noting that although the performance drops slightly, it still significantly outperforms traditional non-pretraining based methods such as BiRNN (MRR=0.29) and SelfAttn (MRR=0.59).

\approach on CodeT5 demonstrates a similar efficacy. The vanilla CodeT5 requires more than 16 hours for fine-tuning and around 3 hours for testing on the Java code search task. By dropping out 40\% input tokens using \approach, the fine-tuning and test time is reduced to around 9 and 1.6 hours, respectively. Meanwhile, \approach (attention) achieves a comparable performance (MRR=0.71) to the original CodeT5 (MRR=0.72) with only a reduction of 1.5\%. 

\approach is also effective in the code summarization task. Taking Java as an example, the vanilla CodeBERT achieves a BLEU-4 score of 18.95 at the expense of 13.82 hours of fine-tuning and 4.8 hours of testing. When the target length is set to 150 tokens (i.e., 60\% of the original length), the fine-tuning time is reduced to about 60\% and the FLOPs drops from 24.328G to only 15.325G. This does not sacrifice accuracy (BLEU-4=17.29).




Comparing the three pruning strategies, \approach with attention shows strength over the two other strategies. For example, \approach with attention-based pruning achieves an MRR of 0.71 in the Java code search task, which is better than dropout (MRR=0.68) and frequency filtering (MRR=0.66). The results suggest that by finer-grained pruning of tokens based on attention weights, \approach can achieve more effective program simplification. \approach on the CodeT5 demonstrates a similar trend in both tasks, where the attention-based pruning achieves better performance (BLEU=19.25) than dropout (BLEU=17.65) and frequency filtering (BLEU=18.74).
We observe that this strength becomes less significant in the Python language. We conjecture that the Python language contains fewer auxiliary tokens (such as brackets), resulting in a more uniform distribution of attention weights over all tokens. Therefore, removing tokens with smaller attention is closer to removing random or frequent tokens in the Python language.



As a noteworthy point, it is inapplicable to test the baseline model SIVAND in our datasets, as it takes a massive amount of time (>30,000 hours according to our estimation) to process more than 1 million functions in the CodeSearchNet benchmarks. This is mainly because the ddmin algorithm used by SIVAND is based on backtracking, where each step the algorithm should invoke the deep learning model for a prediction task.
For this reason, we estimate the time cost of SIVAND by sampling 50 functions from CodeSearchNet. 
Results show that it costs 104 and 77 minutes respectively in the CodeSearchNet (Java) and CodeSearchNet (Python) datasets. 
Based on this observation, we estimate the testing time in the entire dataset. Since the data we sample is small in size, it is insufficient to fine-tune the model to validate the accuracy.

\subsection{Effect of Relative Length (RQ4)}

\begin{figure}
    \subfigure[results on code search]{
    \includegraphics[width=0.47\linewidth, trim=0 0 50 50, clip]{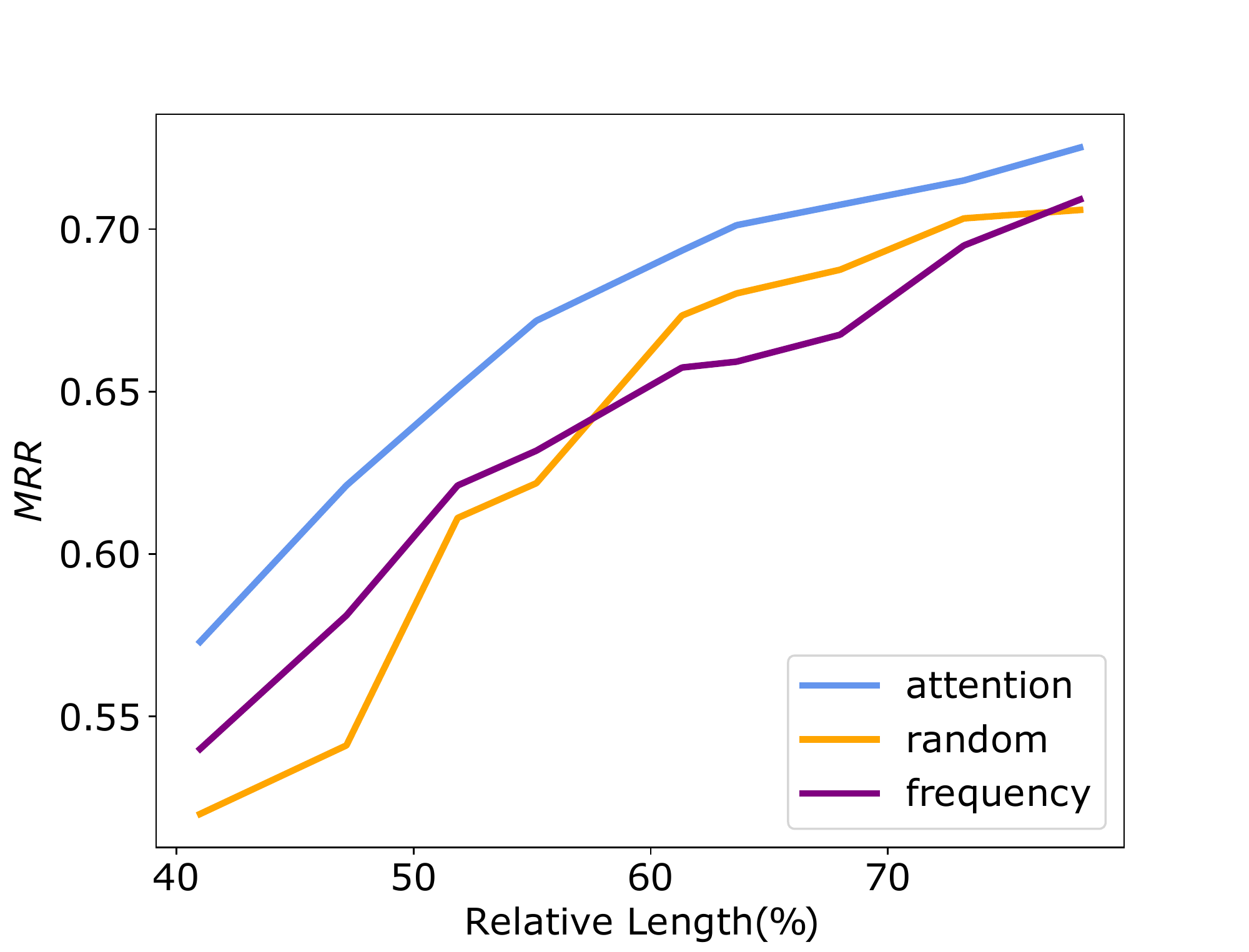}
    \label{fig:downstream_task_reduction}
    }
    \subfigure[results on code summarization]{
    \includegraphics[width=0.47\linewidth,trim=0 0 50 50, clip]{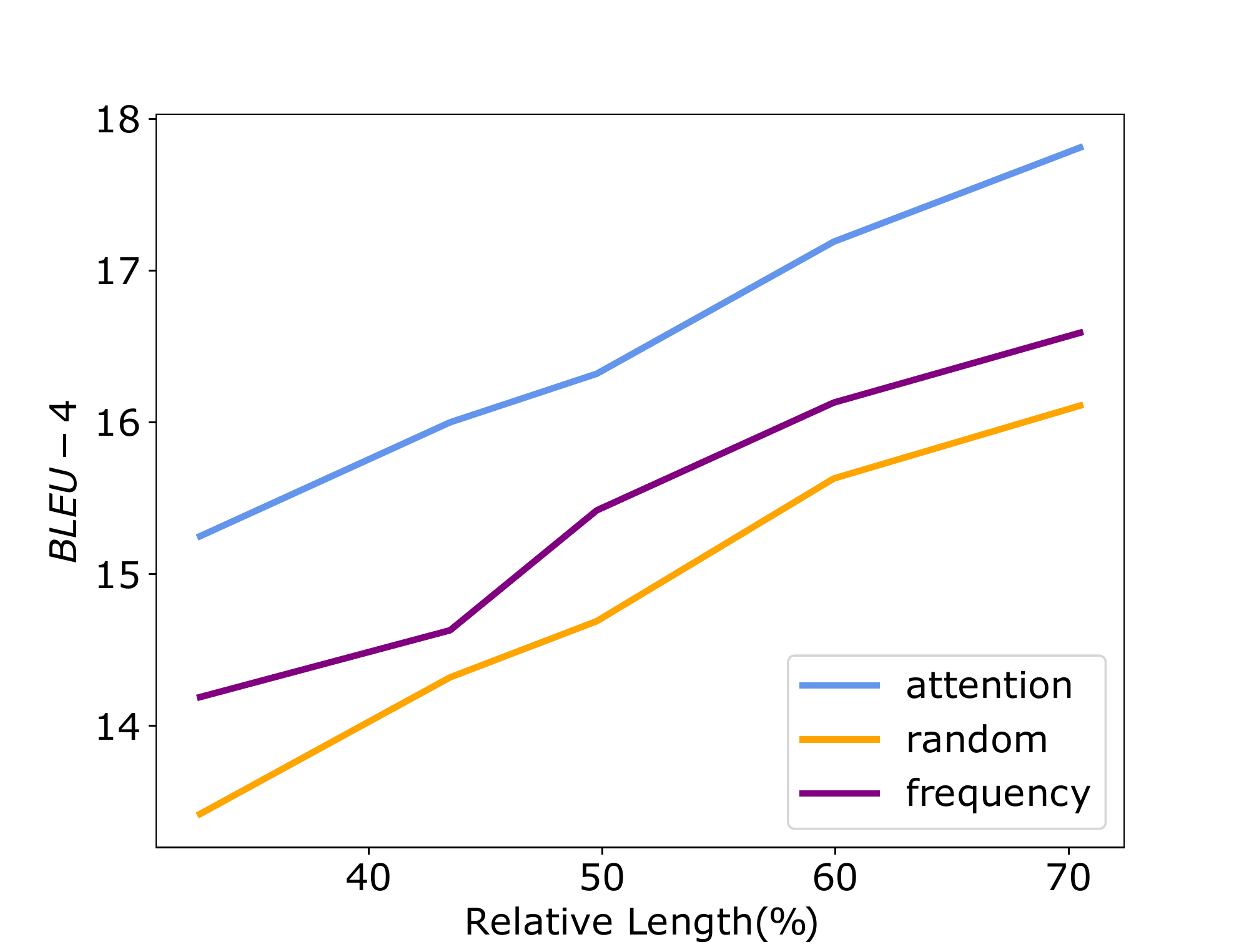}
    \label{fig: token_statement_reduction}
    }
    \caption{Performance under different relative lengths on Java.}
    \label{fig:rq5}
\end{figure}

Figure~\ref{fig:rq5} shows the performance of \approach under different relative lengths and pruning strategies. 
We can see that the performance of all pruning strategies drops sharply as the relative length decreases.
In both two tasks, \approach with attention performs better than the other two pruning strategies under all relative lengths.
When the relative length is above 70\%, the performance is comparable to that of the original CodeBERT but the reduction of computational cost can be modest.
When the relative length drops below 50\%, the performance is deficient since most of the code has been pruned.
A relative length of around 60\% is probably the best trade-off for these two tasks.



\subsection{Effect of Token and Statement Pruning (RQ5)}

\begin{figure}
    \subfigure[results on code search]{
    \includegraphics[width=0.475\linewidth,
    trim=0 0 50 50, clip]{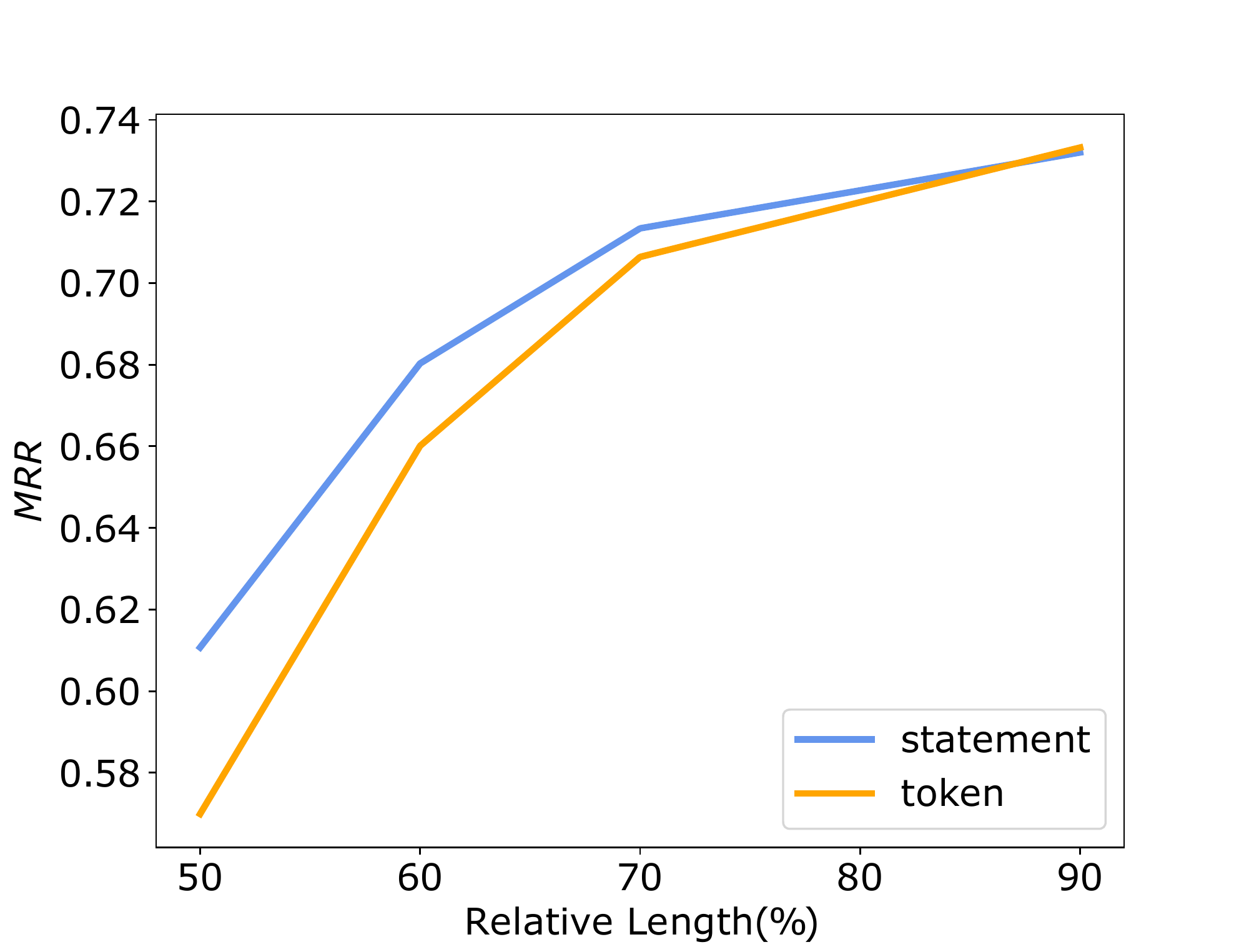}
    \label{fig:tokenAndStatementCodeSearch}
    }
    \subfigure[results on code summarization]{
    \includegraphics[width=0.475\linewidth, trim=0 0 50 50,clip]{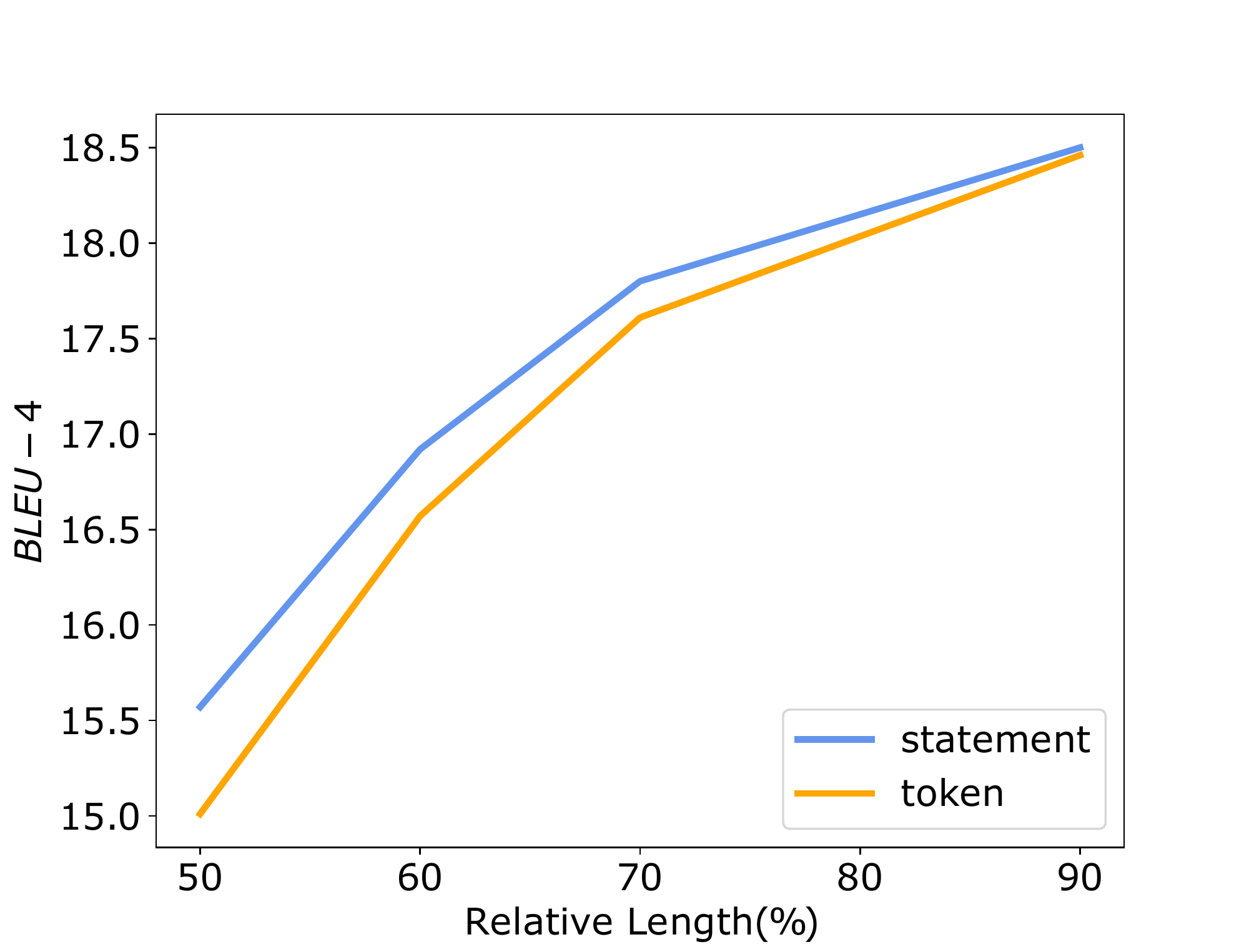}
    \label{fig: tokenAndStatementCodeSummarization}
    }
    \caption{Ablation results of token and statement pruning on Java.}
    \label{fig:rq6}
\end{figure}

Figure~\ref{fig:rq6} shows the performance of \approach with token and statement pruning on the two tasks. 
In order to verify the effect of statement pruning, we only run the statement selection phase in Algorithm 1. To test the effect of token pruning, we drop the lowest attention tokens directly until meeting the target length.
As we can see from the results, a significant drop in performance is observed for both strategies as the relative length decreases.
When 90\% code remains, the MRR scores of the two pruning strategies are close. As the relative length decreases, the score of the token pruning strategy drops much faster than that of the statement pruning strategy. 
When the relative length decreases further, their MRR scores drop dramatically.
The results suggest that statements are more critical than tokens in learning code representations by CodeBERT. 
One possible reason could be that token attention usually measures local information, while statement attention considers the relationship between tokens. Therefore, statement attention could keep more semantic of the functions. 
The code summarization task shares a similar result as the code search task.

%
%

\section{Discussion}

\subsection{How Can Pre-Trained Models Understand Simplified Code?}
One debatable question is how can pre-trained models understand simplified code which is not compilable and runnable. Source code involves two channels of information: formal \& natural~\cite{casalnuovo2020does,chakraborty2022natgen}. It can either be considered as programs for computers to execute 
or 
as a specific natural language for humans to communicate~\cite{hindle2012naturalness}. The former requires the source code to be strictly compilable and runnable, while the latter focuses more on conveying information to humans. 
We believe that CodeBERT considers more about the natural channel of source code~\cite{hindle2012naturalness,chakraborty2022natgen}. As such, it pays more attention to keywords without necessarily having to know how the program actually behaves during execution. 

Our empirical findings are in agreement with a previous human study on how programmers summarize source code~\cite{rodeghero2014improving}, which shows that developers usually start by reading the function name. They may skip some structural information such as \texttt{For} and \texttt{If} conditions while paying special attention to literal information such as method invocations and variable names, which literally show the intention of the code~\cite{rodeghero2014improving}. 
The pruning algorithms described in our paper make use of this finding and simplify source code by only selecting critical information (e.g., reflected by the attentions of CodeBERT) in the source code. Therefore, they do not have much impact on the understanding of source code.

\subsection{What users may adopt the techniques?}
One debatable question is what users may adopt the technique since it has a precision loss.
From the results in Table~\ref{tab:codesearch_result} and \ref{tab:codesummarize_result}, although the performance drops slightly, it still significantly outperforms traditional non-pretraining based methods such as BiRNN and SelfAttn.  
Meanwhile, \approach shows an advantage of computational efficiency (e.g., saving 40\% of fine-tuning time). Hence, \approach provides a practical option for users with limited computational resources but still want quick leverage of large pre-trained models.

\subsection{The role of the 0-1 knapsack formulation}
One may question whether we really need to transform the problem into the 0-1 knapsack.
In our preliminary studies, simply pruning statements often results in sequences that are either too long or too short. Token pruning provides a finer-grained pruning to the input program. This amounts to selecting statements and tokens to obtain the maximum attention while satisfying the length restriction. Hence, we formulate it as a 0-1 knapsack problem and solve it using a greedy strategy.
According to the experimental results in Figure~\ref{fig:rq6}, a simple statement pruning archives an MRR of around 0.68 on Java code search when the relative length is 60\%, which is evidently lower than that of \approach (MRR=0.71). The same comparison can be observed on the Java code summarization task (i.e., 16.8 vs. 17.3 in terms of BLEU). This suggests the effectiveness of the 0-1 knapsack formulation.

\subsection{Implications for Future Research}
Through our empirical study, we find that CodeBERT does not recognize grammar and structural information very well.
For example, the \texttt{if\;condition}, \texttt{for\;condition}, and \texttt{while\;condition} are three typical statements representing code structures, but they are rarely considered by pre-trained models when learning program representations.
This motivates future research on better incorporating code structures into pre-trained code models, for example, converting structural symbols (e.g., the bracket \texttt{\}} ) to a more verbal style (e.g., `the end of a for statement') before feeding into pre-trained models. 

Besides \approach, there is a broader class of techniques, including model compression~\cite{shen2020q-bert}, model distillation~\cite{sanh2019distilbert}, and data reduction~\cite{ye2021tr-bert}, which aim to reduce the computational complexity of large pre-trained models at the cost of a slight decrease in performance.
\approach provides an easy, alternative solution by only pruning input programs. We believe this is more practical for developers as they only need to process the data without modifying the model. 
In the future, we can investigate more techniques such as model compression and distillation.

\subsection{Threats to Validity}
\emph{Internal Threats.} 
When calculating the attention weights for statements, we encounter the out-of-vocabulary (OOV) issue. We resolve this issue by excluding rare categories of statements in our experiments. Nevertheless, some categories of statements could still be necessary though having a low quantity in the corpus. 
Furthermore, the raw data provided by CodeSearchNet is in the format of plain text. We parse statements using text processing rules designed by ourselves. However, there could be irregular patterns that our parser can not recognize. This could cause noise in our dataset and affect the results of our research.

\emph{External Threats.} 
We have just conducted our experiments in Java and Python. Though the conclusions from both languages are similar, other programming languages such as LISP and Erlang could have different attention patterns. 
In addition, \approach is evaluated in two downstream tasks: code search and code summarization.
However, these tasks rely much on the literal information provided in the source code. Thus, it remains to be verified whether or not the proposed simplification algorithm is applicable to other software engineering tasks.

\section{Related Work}

\subsection{Understanding Pre-trained Models of Code}


Besides our work, there have been other studies that also try to explain the mechanisms of pre-trained models for code~\cite{ahmad2021plbart,mastropaolo2021t5code,rogers2020bertology,paltenghi2021thinking}. 
~\citet{karmakar2021pre} applied four probing tasks on pre-trained code models to investigate whether pre-trained models can learn different aspects of source code such as syntactic, structural, surface-level, and semantic information.
Different from our work, they only empirically study the whole pre-trained model, whereas we focus on more specific (critical tokens and statements) knowledge learned by pre-trained models. 

There have also been many works that investigate the attention weights of pre-trained models for source code. 
For example, ~\citet{wan2022they} performed a structural analysis of pre-trained language models for source code. They analyzed the self-attention weights and found that the Transformer attention can capture high-level structural information of source code.
AutoFocus~\cite{bui2019autofocus} aims to reveal the most relevant part of the code to programmers. 
They measure the relevance of statements using attention weights from a GGNN~\cite{allamanis2017learning}.
Different from our approach, they capture structural knowledge learned by pre-trained models, while we dig more into critical statements and tokens learned by pre-trained models. Furthermore, we propose simplifying programs for pre-trained language models based on the findings.

\subsection{Program Simplification}

Program simplification has gained increasing attention recently. The state-of-the-art methods such as SIVAND~\cite{rabin2021simplification} and P2IM~\cite{suneja2021probing} a based on the delta debugging prototype~\cite{zeller2002simplifying}. The delta debugging mechanism requires an input code snippet and an auxiliary deep learning model such as the code2vec~\cite{alon2019code2vec}.
The deep learning model takes as input the code snippet and splits it into fragments. Each fragment is then taken as input to the neural network model to perform testing tasks such as method name prediction and misused variables detection.
If a fragment obtains a satisfying score, the program will split it further. The process continues until the performance of the subset does not satisfy the target score.
Finally, the algorithm produces the smallest code snippet that satisfies the objective of the deep model.

Compared to \approach, their methods are computationally inefficient because they need to run an auxiliary deep learning model and evaluate the performance at each iteration.
For example, it costs hundreds of hours for SIVAND to process 10,000 functions, while \approach can complete it within two minutes.

\section{Conclusion}
This paper empirically analyzes the critical information learned by CodeBERT, including important tokens and statements in both Java and Python. Our results show that CodeBERT focuses on keywords and data-relevant statements such as method signatures and return statements.
Based on our empirical findings, we propose a novel approach named \approach for lightweight leverage of pre-trained models. \approach simplifies the input code to a target length by selecting important statements and tokens based on their attention weights using a 0-1 Knapsack algorithm.
Experiments on two tasks have shown that \approach provides comparable results as CodeBERT, with the advantage of computational efficiency.

In the future, we will consider more aspects that CodeBERT learns about code, such as syntax rules and semantic relations, to further reduce the size of source code for pre-trained models. 
We will also investigate model compression~\cite{cheng2017survey,cheng2018model} and distillation~\cite{sanh2019distilbert} techniques to further reduce the size of pre-trained programming language models.

Our source code and experimental data are publicly available at \href{https://github.com/zhangzwwww/DietCode}{https://github.com/zhangzwwww/DietCode}

	
\section{Acknowledge}
 This work was sponsored by NSFC No. 62102244, CCF-Tencent Open Research Fund (RAGR20220129), and CCF-Baidu Open Fund (NO. 2021PP15002000). 
 
\bibliographystyle{ACM-Reference-Format}
\balance
\bibliography{references}

\end{document}